# The effect of Sn intercalation on the superconducting properties of 2H-NbSe$_2$

Subham Naik[1], Gopal K. Pradhan[2], Shwetha G. Bhat[3], Bhaskar Chandra Behera[2], P. S. Anil Kumar[3], Saroj L Samal[1*], DSamal[2, 4*]

[1]Department of Chemistry, NIT Rourkela, Rourkela-769008, Odisha, INDIA

[2]Institute of Physics, Bhubaneswar, Bhubaneswar-751005, INDIA

[3]Department of Physics, IISc Bangalore, Bangalore, INDIA

[4]Homi Bhabha National Institute, Anushakti Nagar, Mumbai 400085, INDIA

2H-NbSe$_2$ is known to be an archetype layered transitional metal dichalcogenide superconductor with a superconducting transition temperature of 7.3 K.In this article, we investigate the influence of Sn intercalation on superconducting properties of 2H-NbSe$_2$. Sn being nonmagnetic and having no outer shell d-electronsunlike transition metals, one naively would presume that its effect on superconducting properties will be very marginal. However, our magnetic and transport studies reveal a significant reduction of both superconducting transition temperature and upper critical field [$T_c$ and $B_{C2}$ (0)] upon Sn intercalation. With a mere 4 mole% Sn intercalation, it is observed that $T_c$ and $B_{C2}$ (0) get suppressed by ~3.5K and 3T, respectively. Werthamer-Helfand-Hohenberg (WHH) analysis of magneto-transport data is performed to estimate$B_{C2}$ (0). From the low temperature Raman scattering data in the normal phase of intercalated NbSe$_2$, it is inferred that the suppression of superconductivity cannot be ascribed to strengthening of charge density wave (CDW)ordering. The effects such as electron-doping induced Fermi surface change and/or disorder scattering upon intercalation are speculated to be at play for the observed phenomena.

# INTRODUCTION

Quasi-two-dimensional layered transition metal dichalcogenides (TMDs) have beenthesubject of intense research owing to their rich electronic properties resulting from lower dimensionality. These compounds share the $MX_2$ formula, where M is a transition metal (M = Ti, Zr, Hf, V, Nb, Ta, Mo, W or Re), and X is a chalcogen (X = S, Se, or Te). The underlying structures are made from stacking X-M-X layers in repeating patterns with inter-layer Van der Waals bonding. The weak Van der Waals interlayer bonding between hexagonal layers of octahedral or trigonal prismatic TMD building blocks allows many polytypes to form.Many of the members in this family exhibit the coexistence/competition between charge density wave (CDW) and superconductivity; where the superconductivity is found to emerge in the vicinity of CDW phase [1, 2].The overall electronic phase diagram of these layeredsuperconductors arestrikingly similar to those found in the high-$T_c$cuprates, the organic layered superconductors, and the iron-based pnictides[3].One of the earliest layered TMD materials known to superconduct is 2H-NbSe$_2$ with a transitiontemperature ($T_c$) ~7.3K, which is significantly higher than its compatriot superconductors known, where $T_c$ is commonly in the range of 2 - 4 K. It hosts a quasi-two-dimensional charge density wave (CDW) with CDW critical temperature ($T_{CDW}$) ~ 33 K that coexists at local level with superconductivity and also has a strong superconducting gap anisotropy [4]. It is not yet clear whether the observed superconducting gap anisotropy in 2H-NbSe$_2$ is a result of there being different gaps on different Fermi surface sheets, or it originates elsewhere. It is also found that 2H-NbSe$_2$ is a multiband superconductor with some similarities to that of MgB$_2$ superconductor[5,6]. Despite the fact that 2H-NbSe$_2$ has been considered as a conventional superconductor, the delicate balance that threads superconductivity, CDW, strong gap anisotropy and its multiband character are still elusive and under active discussion [6 – 9].The electronic properties of 2H-NbSe$_2$ are greatly influenced by applied pressure, doping, intercalation and layer thickness[7, 8, 10–12]. Hydrostatic pressure has been shown to be increasing $T_c$ while decreasing$T_{CDW}$, and an increase of the effective dimensionality of electronic structure[10]. It is well known that for *s*-wave pairing isotropic superconductors, scattering by nonmagnetic impurities has no effect on $T_c$[13,14]. However, the $T_c$ for unconventional superconductors is particularly sensitive to scattering by nonmagnetic impurities/defects that can destroy the pairing strength[15]. Besides, the nature of relationship between the interlayer distance and $T_c$ upon intercalation is also a matter of fundamental importance in the understanding of superconductivity of the layered compounds. Theoretical investigation on the effect of intercalation on the $T_c$ of $NbSe_2$ predict that the donor-type intercalant lowers the $T_c$, whereas, the acceptor-type intercalant raises the $T_c$, as long as there is no lattice instability[16]. It is to be noted that very recent investigation on Cu intercalated 2H-$NbSe_2$ showed an unusual S-shaped suppression of superconductivity [11].

In this article, we report the results on the influence of Sn intercalation on superconducting properties of 2H-$NbSe_2$. Sn intercalation in the doping range ($0 \leq x \leq 0.04$) is made to form $Sn_xNbSe_2$ as shown in Fig.1 (a) that retains the 2H structure without any impurity phase.For samples having higher concentration of Snshowing the signature for $SnSe_2$ impurity phase are excluded in this study.Detailed structural, spectroscopic, magnetic and transport studies were carried out to correlate the superconducting property with Sn intercalation. We find that both $T_c$ and $B_{C2}$ (0)decrease significantly in Sn-intercalated $NbSe_2$ as compared to that of parent phase. This is in line with the results reported by Luo*et al*.[17]on Cu intercalated $NbSe_2$ though Cu and Sn are electronically different entity. Our preliminary analysis suggests that the suppression cannot be ascribed to strengthening of CDW or increasing of inter-layer distance. However, electron-doping induced Fermi surface change in $NbSe_2$ layer upon Sn intercalation and/or scattering by disorder effect turns out to be plausible reasons for the observed phenomena.

**EXPERIMENTAL**

Polycrystalline samples of$Sn_xNbSe_2$ were synthesized by high temperature reaction using solid state method.Stoichiometric amount of high pure niobium powder (Sigma Aldrich; 99.9%),selenium powder (Sigma Aldrich; 99.5%) and tin shots (Sigma aldrich; 99.8%) were weighed and pelletized using hydraulic press. The prepared pellets were evacuated in a silica tube (~$10^{-3}$Torr), sealed and then heated to 700$^o$C for 24 hrs followed by quenching to room temperature. The product formed was further ground, palletized, and sealed in a silica tube under vacuum and reheated at the above temperature. All the reactants were handled inside a glove box filled with nitrogen. Phase purity were determined by powder X-ray diffraction (XRD) on a RigakuUltima IV multipurpose X-ray diffractometer using Cu K$\alpha$ ($\lambda$=1.54Å) source.The lattice parameter refinements were accomplished using WinXPow[18].The field emission scanning electron micrographs (FESEM) of the $Sn_xNbSe_2$ compounds were recorded on a NOVA NANOSEM 450 field emission scanning electron microscope. FESEM of $Nb_{1-x}Sn_xSe_2$ shows layered morphology (see supplementary material Figure. S1). Measurements of the temperature dependent electrical resistivity and magnetization down to 2K of $Sn_xNbSe_2$ were performed using physical property measurement system (PPMS) and superconducting quantum interference device (SQUID) respectively. The four-probe geometry was used for the resistivity measurement. Raman scattering measurements were performed using Jobin-Yvon T64000 triple monochromator based Raman system in the backscattering micro-Raman configuration. The 514.5 nm line of an $Ar^+$ ion laser was used as an excitation source. The laser power was kept low to avoid sample damage and heating.

**Results and Discussion**

$NbSe_2$ is a layered compound in which, Nb is bonded to six selenium atoms in trigonal prismatic geometry and each Se atom bonded to three niobium atoms (Figure 1(a). The intra and inter layer thickness are 3.316 Å and 2.965 Å, respectively. So the layered compound offers space for the intercalation of smaller cations. The PXRD pattern for $NbSe_2$ and Sn-interacted $NbSe_2$ in Figure.1 indicates that all the reflections can be indexed with 2H-$NbSe_2$ structure without any extrinsic phases. It is observed that peak positions shift towards lower angle with increase in Sn concentration (left inset in Figure.1(b)). The refined lattice parameters of $Sn_xNbSe_2$ are shown as an inset (right) in Figure.1 (b). It is observed that the *c*-lattice parameter gradually increases with Sn concentration up to $x = 0.04$. However, a very subtle change (in the third decimal place) is observed for the '*a*' lattice parameter with Sn concentration. As we know that the $r_{Sn4+}$ (0.69Å in CN =6) is very much similar to that of $r_{Nb4+}$ (0.69Å in CN =6), so the increase in the *c*-lattice parameter could be due to Sn intercalation between the $NbSe_2$ layers.

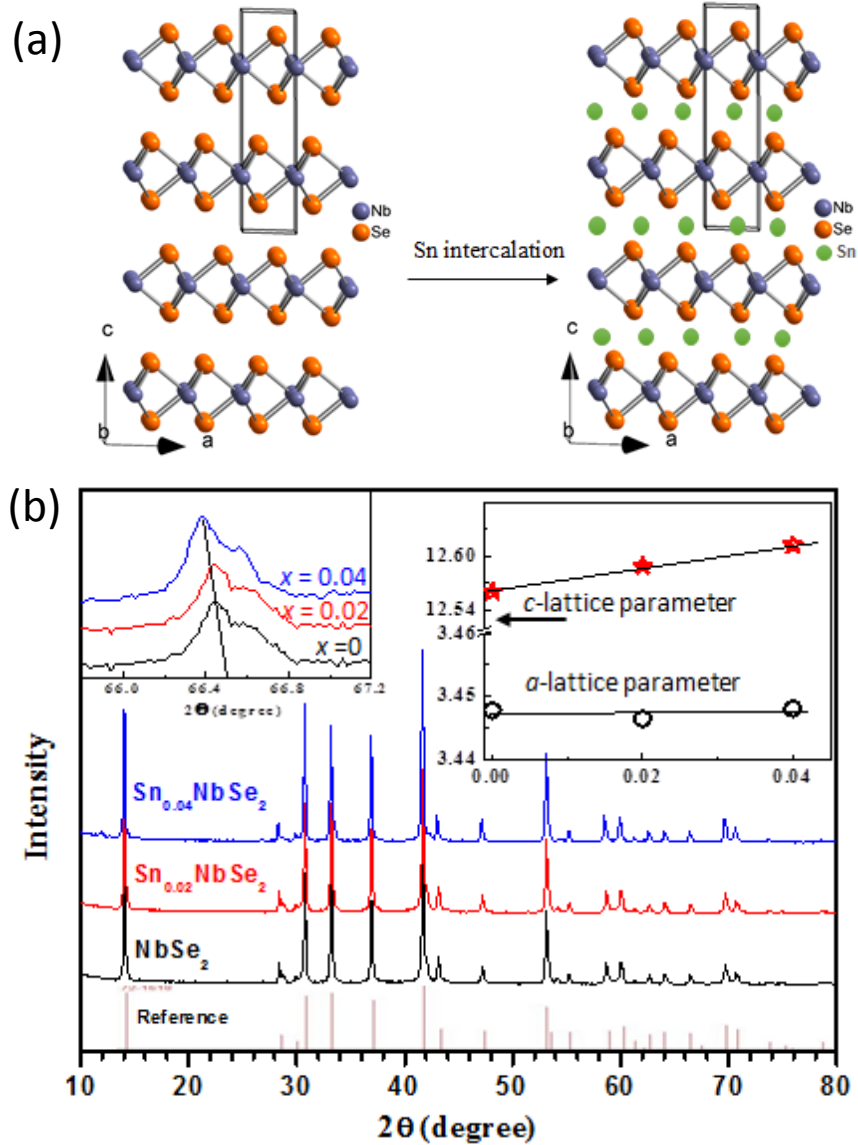

**Fig.1** (a) Schematic for Sn intercalation of 2H-NbSe$_2$, (b) shows the powder X-ray diffraction patterns for Sn$_x$NbSe$_2$ ($0 \leq x \leq 0.04$). The results show formation of a single-phase solid solution. The left inset in 1(b) shows the 2θ-region where the peak shifts towards lower angle with increase in $x$ is clearly seen. The right inset in 1(b) shows variation of $a$ and $c$ lattice parameters.

To explore how the superconducting properties evolve with Sn intercalation in Sn$_x$NbSe$_2$, we first investigated the temperature dependent magnetization behavior. Fig.2 (a) shows the typical M-T curve signaling the superconducting (diamagnetic) state. However, the onset of the diamagnetism shifts systematically to lower temperatures with increasing $x$ as shown in the inset. This suggests that Sn-intercalation may be an electronically disruptive for the superconductivity in NbSe$_2$. Moreoever, we observe that the absolute value of diamagnetic response decreases in the superconducting state upon Sn intercalation, possibly because of the reduction in supercondcuting volume fraction [19].

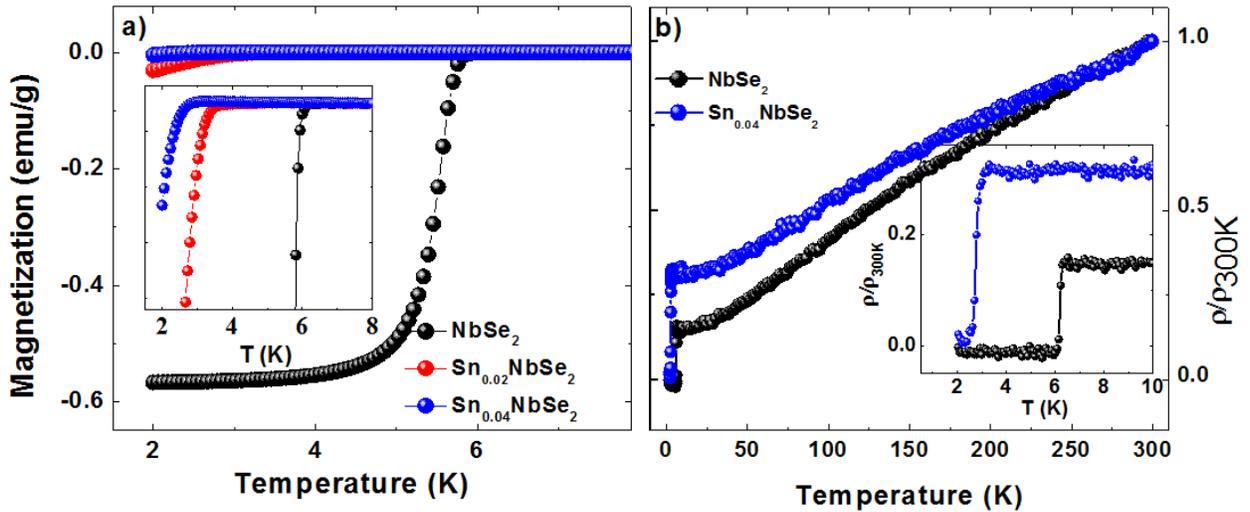

**Fig.2** (a) Temperature dependent zero field cooled (dc field 50 Oe) magnetization for Sn$_x$NbSe$_2$ (0 ≤ $x$ ≤0.04) revealing superconducting transition. The inset shows the magnified view of superconducting transition. (b) Normalized temperature dependent resistivity for Sn$_x$NbSe$_2$ with $x$=0 and $x$=0.04.The inset shows the magnified view of superconducting transition.

The temperature dependent resistivity of two representative Nb$_{1-x}$Sn$_x$Se$_2$ samples with $x$=0 and $x$=0.04 is shown in Fig. 2(b). Both the samples show metallic temperature dependence (d$\rho$/d$T$> 0) in the normal state. However, the residual resistivity ratio (RRR) of NbSe$_2$ is, $\rho_{300K}/\rho_{10K}$ = 6.72 and that of Sn$_{0.04}$NbSe$_2$ is, $\rho_{300K}/\rho_{10K}$ =3.13.This indicates thatSn$_{0.04}$NbSe$_2$has more electronic scattering than NbSe$_2$. Generally, high quality single crystals and ultrathin layer of NbSe$_2$ with high RRR show a broad shoulder/hump around 30K in the $\rho$–$T$ curve revealing the emergence of hidden CDW [20,21]. In our case, we do not see such effect, which could get masked by scattering from grain boundaries and other extrinsic defects due topolycrystalline nature of the sample. Although careful interpretation of resistivity necessitates the use of data obtained on single crystals, consideration of the data on polycrystalline samples can provide some basic insights. At low temperatures (see Fig. 2 (b)), a sharp drop of $\rho(T)$ is observed in both the samples signifying the superconducting transition. The inset in Fig. 2(b) shows the magnifiedview of the superconducting transitions. A significant reduction in the superconducting $T_c$is observed for $x$=0.04 as compared to that parent $NbSe_2$ (the onset of superconducting transition for $NbSe_2$ and $Sn_{0.04}NbSe_2$ is found to occur at ~ 6.3 and 3.15 K, respectively). The suppression of superconducting transition observed from transport study are in good agreement with the magnetization data. The origin of reduced superconductivity in Sn intercalated $NbSe_2$ is difficult to singularize in the present study; however few rational reasons could be identified. Hydrostatic pressure has been shown to lead to an elevation of $T_c$ in $NbSe_2$[7,22,23]. It is to be mentioned that the increment of $T_c$ in $NbSe_2$ has been reported to be ~0.25-0.86 K/GPa under the application of hydrostatic pressure.[7,22]As a first approximation, this could be due to enhanced interlayer interaction resulting from the reduction in the interlayer distance. In contrast to hydrostatic pressure effect, the intercalation gives rise to an increase of interlayer distance and possibly could have detrimental effect on superconductivity. In our case, we find an increase in $c$-axis lattice parameter by a relatively smaller value of 0.4 % for $x$=0.4, and hence we believe that it cannot solely be accounted for the observed effect. We, rather conjecture that doping induced effect as well as disorder effect introduced upon Sn intercalation could be at play for the reduction in superconductivity.

It may be recalled that superconducting $T_c$ and $T_{CDW}$ shows an anti-correlation upon application of pressure or when one reduces the number of layers in $NbSe_2$[10,24,25]. In the absence of signatures of CDW transition from the ρ–T data due to polycrystalline nature of the samples, we turned our attention to Raman scattering to see the trend in$T_{CDW}$ if any, with Sn intercalation. We thus carried out Raman spectroscopic measurements in the normal phase of $NbSe_2$ and$Sn_{0.04}NbSe_2$in the temperature range 300 – 40 K.

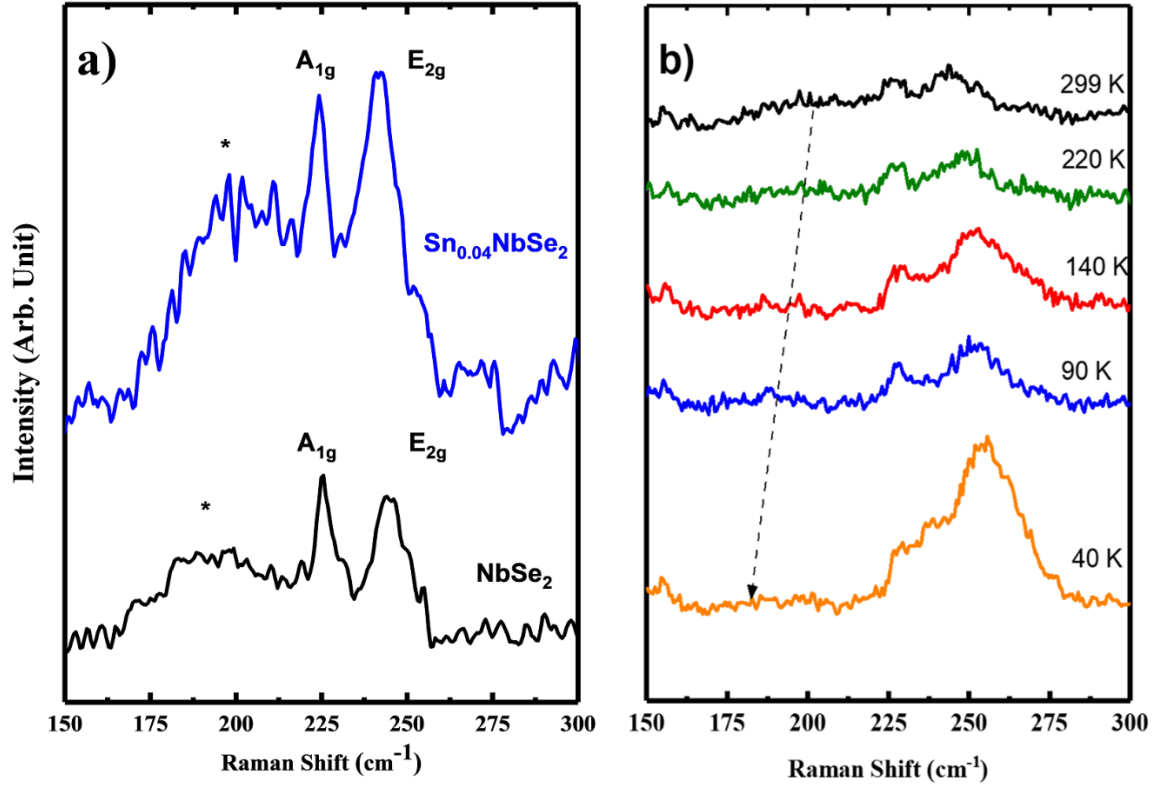

**Fig. 3** a) Raman spectra of NbSe$_2$ and Sn$_{0.04}$NbSe$_2$. The broad feature(marked as *) around 190 cm$^{-1}$ peak involving a second-order scatteringprocess of two phonons. Disappearance of this mode is identified with CDW transition[26]. b) Temperature evolution of the Raman spectra of Sn$_{0.04}$NbSe$_2$ at representative temperatures. The broad featuresurvives until 40 K achieved in this experiment. The dashed arrow showing the red shift of the two-phonon mode is a guide to the eye.

Figure 3a) shows the ambient Raman spectra of NbSe$_2$ and Sn$_{0.04}$NbSe$_2$. The prominent features observed below 300 cm$^{-1}$ include the well resolved in-plane phonon E$_{2g}$ modes and an out-of-plane A$_{1g}$ phonon mode at at∼245 cm$^{-1}$ and225 cm$^{-1}$, respectively[27,28]. Apart from these well-known Raman active normal modes, the normal phase NbSe$_2$ displays an anomalous two-phonon broad feature (marked as * in Fig. 3 a) of A$_{1g}$ symmetry around 190 cm$^{-1}$. This is assigned as a soft mode, which involves a second-order scattering process of two phonons of frequency ω$_0$ at wavevector ∼(2/3)Γ-M[27,29]. This mode, which softens with temperature, freezes below T$_{CDW}$ (33 K) for NbSe$_2$ is usually taken as a marker for the CDW transition[27,28]. Figure 3 (b) summarizes the temperature dependence of the Raman spectra for

4% Sn intercalated NbSe$_2$, from 40 to 300 K. The temperature behavior of the soft mode agrees very well with that of the bulk NbSe$_2$, namely the mode redshifts with decreasing temperature. It is interesting to note that the soft mode feature survives up to 40 K achieved in this experiment, indicating the absence of CDW down to 40K. Thus, nature of the CDW order and its relationship to suppressed superconductivity in intercalated NbSe$_2$ can possibly be ruled out. However, other effects such as the modified structure of the Fermi surface and/or disorder induced scattering as mentioned earlier must be considered to understand the observed T$_c$ suppression.

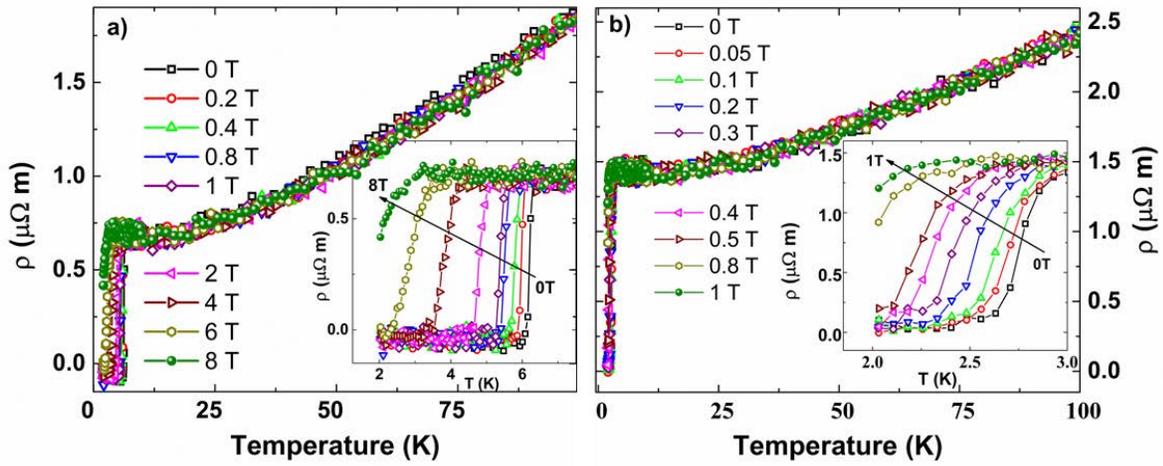

**Fig.4** ρ-T plots for (a) NbSe$_2$ and (b) Sn$_{0.04}$NbSe$_2$ under various applied magneticfield. The inset shows the magnified view of superconducting transition.

In Fig.4, we show temperature dependent resistive transition curves under the application of differentmagnetic fields (the insets in in Fig.4 (a) and (b) show the magnified view of the superconducting transitions). As expected, the superconducting transition gets suppressed with the applied magnetic field. To compare how $B_{C2}$–$T_c$ phase evolves with Sn-intercalation, we employed $\rho$-$T$ plots under applied magnetic field to extract the same. The temperature dependent upper critical field for type-II superconductors is well described by Werthamer-Helfand-Hohenberg (WHH) formalism [30–32].

We use a simplified WHH formalism, which excludes spin-paramagnetic and spin-orbit effects. The WHH equation considering only the orbital effects in the dirty limit is given by:

$$\ln\left(\frac{1}{t}\right) = \psi\left(\frac{1}{2} + \frac{h}{2t}\right) - \psi\left(\frac{1}{2}\right) \quad (1)$$

Where $\psi$ is the Digamma function and $t = T/T_c$ is the reduced temperature and "$h$" is given by the following relation:

$$h = \frac{4B_{C2}}{\pi^2(-dB_{C2}/dT)_{T=T_c}}$$

As T→0, equation (1) reduces to $B_{C2}(0) = 0.693 T_C (dB_{C2}/dT)_{T=T_c}$ (2)

Both the above equations (1 & 2) depend on the term $dB_{C2}/dT$, which is a crucial parameter for the determination of $B_{C2}(0)$. In practice, $B_{C2}(0)$ can be obtained either by numerically solving equation (1) that fits the data or directly from equation (2). Using 50% normal state resistance criterion as the marking for $T_c$, we obtained $B_{C2}$–$T_c$ phase diagram as shown in Figure 5. It is noteworthy that $B_{C2}(0)$ for $Sn_{0.04}NbSe_2$ (1.75 T) is significantly reduced as compared to that of $NbSe_2$ (5.15 T). The upper critical field for bulk single crystalline $NbSe_2$ is reported to be $B_{C2\perp}(0) \sim 4$ T and $B_{C2//}(0) \sim 12$ T respectively[5]. Recent work by Luo *et al* [17] on $NbSe_2$ polycrystalline sample with $T_c \sim 7.16$K estimated $B_{C2}(0)$ to be of ~9.7 T. The lower $B_{C2}(0)$ observed in our case could partly be related to the reduced critical temperature. This is because based on Ginzburg-Landau theory, upper critical field $B_{c2}(T)$ related to coherence length $\xi(T)$ follows a relation: $B_{c2}(T) = \Phi_0/2\pi\xi^2(T)$, with $\xi^2(T) = \xi_0^2 \frac{1}{1-T/T_c}$ and $\Phi_0$ being the superconducting flux quantum[33].

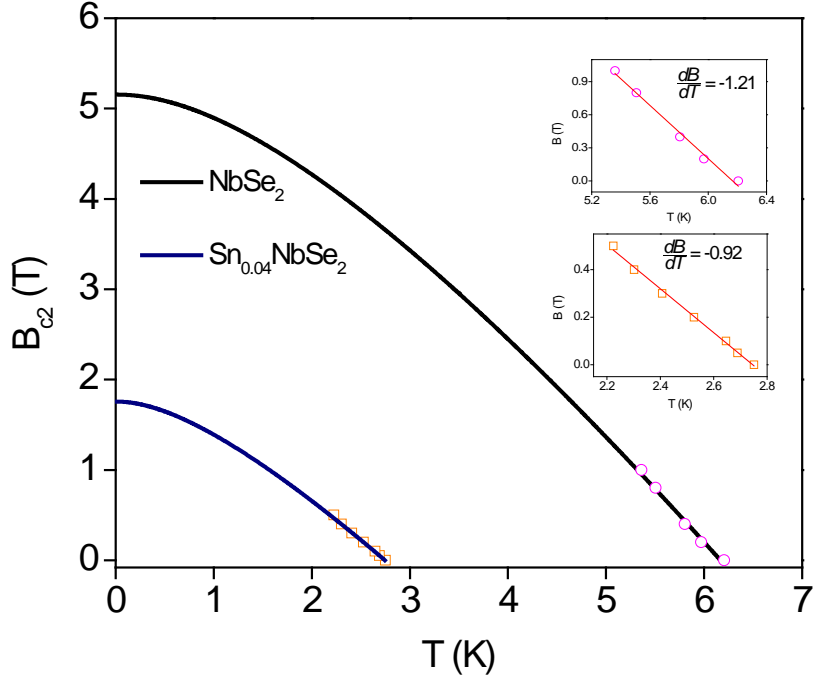

**Fig.5** $B_{C2}$–$T_c$ phase diagrams for NbSe$_2$ and Sn$_{0.04}$NbSe$_2$ extracted using WHH formalism. The inset shows the slope dB/dT in the vicinity of T$_c$.

**Conclusion**

In summary, we have presented results on the variation of superconducting properties upon Sn intercalation upto 4 mole% into 2H-NbSe$_2$. By performing magnetic and magnetotransport measurements, we demonstrate that that both T$_c$ and $B_{C2}(0)$ decrease significantly in Sn-intercalated NbSe$_2$ as compared to that of parent NbSe$_2$. We used a simple WHH formalism to estimate $B_{C2}(0)$ using 50% normal state resistance criterion. With 4 mole% Sn intercalation, it is observed that T$_c$ and $B_{C2}(0)$ get suppressed by ~3.5K and 3T respectively. A relatively small (0.4 % for $x = 0.04$) increase in c-axis lattice parameter measured by XRD and hardly any evidence of strengthening of CDW ordering from low temperature Raman scattering data indicates that the observed decrease in T$_c$ could be due to other effects. We believe that doping of NbSe$_2$ layer upon Sn intercalation and/or non-magnetic disorder effect introduced by Sn intercalation can be at play for such reduced effect. A detailed theoretical modelling and extensive experimental investigation in this regard is warranted for future study.

**ACKNOWLEDGMENTS**

DS acknowledges the financial support from Max-Planck Society through Max Planck Partner Group and thanks S Verma, Institute of Physics, Bhubaneswar for extending the Raman experiment facilities. DS also thanks A Gaurav for the inputs in data analysis. SLS acknowledges financial support from Department of Science and Technology (DST) Science and Engineering Research Board (SERB) (YSS/2015/000038), Government of India. SGB thanks DST, India for the INSPIRE faculty award.

*dsamal@iopb.res.in
* samalsaroj@nitrkl.ac.in